\documentclass[aps,prl,amssymb,twocolumn,amsmath,superscriptaddress,showpacs,10pt]{revtex4}

\usepackage{graphicx}
\usepackage{dcolumn}
\usepackage{bm}
\usepackage{color}
\usepackage{epstopdf}

\begin{document}

\newcommand{\beq}{\begin{equation}}
\newcommand{\eeq}{\end{equation}}
\newcommand{\barr}{\begin{eqnarray}}
\newcommand{\earr}{\end{eqnarray}}

\newcommand{\RED}[1]{\textrm{\color{red}#1}}
\newcommand{\rev}[1]{{\color{red}#1}}

\newcommand{\REV}[1]{{\color{red}#1}}
\newcommand{\BLUE}[1]{{\color{blue}[[Saverio: #1]]}}
\newcommand{\GREEN}[1]{{\color{green}[[PAOLO: #1]]}}

\newcommand{\andy}[1]{ }
\newcommand{\bmsub}[1]{\mbox{\boldmath\scriptsize $#1$}}

\def\R{\mathbb{R}}
\def\C{\mathbb{C}}
\def\bra#1{\langle #1 |}
\def\ket#1{| #1 \rangle}
\def\sinc{\mathop{\text{sinc}}\nolimits}
\def\cV{\mathcal{V}}
\def\cH{\mathcal{H}}
\def\cT{\mathcal{T}}
\def\cM{\mathcal{M}}
\def\cN{\mathcal{N}}
\def\CW{\mathcal{W}}
\def\cA{\mathcal{A}}
\def\e{\mathrm{e}}
\def\ii{\mathrm{i}}
\def\d{\mathrm{d}}
\renewcommand{\Re}{\mathop{\text{Re}}\nolimits}
\newcommand{\tr}{\mathop{\text{Tr}}\nolimits}

\newcommand{\MN}{M_N(\mathbb{C})}
\newcommand{\Mk}{M_k(\mathbb{C})}
\newcommand{\id}{\mbox{id}}
\newcommand{\ot}{{\,\otimes\,}}
\newcommand{{\Cd}}{{\mathbb{C}^d}}
\newcommand{\sbsigma}{{\mbox{\scriptsize \boldmath $\sigma$}}}
\newcommand{\sbalpha}{{\mbox{\scriptsize \boldmath $\alpha$}}}
\newcommand{\sbbeta}{{\mbox{\scriptsize \boldmath $\beta$}}}
\newcommand{\bsigma}{{\mbox{ \boldmath $\sigma$}}}
\newcommand{\balpha}{{\mbox{ \boldmath $\alpha$}}}
\newcommand{\bbeta}{{\mbox{ \boldmath $\beta$}}}
\newcommand{\bmu}{{\mbox{ \boldmath $\mu$}}}
\newcommand{\bnu}{{\mbox{ \boldmath $\nu$}}}
\newcommand{\ba}{{\mbox{ \boldmath $a$}}}
\newcommand{\bb}{{\mbox{ \boldmath $b$}}}
\newcommand{\sba}{{\mbox{\scriptsize \boldmath $a$}}}
\newcommand{\sbb}{{\mbox{\scriptsize \boldmath $b$}}}
\newcommand{\sbmu}{{\mbox{\scriptsize \boldmath $\mu$}}}
\newcommand{\sbnu}{{\mbox{\scriptsize \boldmath $\nu$}}}
\def\oper{{\mathchoice{\rm 1\mskip-4mu l}{\rm 1\mskip-4mu l}%
{\rm 1\mskip-4.5mu l}{\rm 1\mskip-5mu l}}}
\def\<{\langle}
\def\>{\rangle}
\newtheorem{theorem}{Theorem}
\newtheorem{definition}{Definition}
\newtheorem{remark}{Remark}

\title{The observables of a dissipative quantum system }

\author{Dariusz Chru\'sci\'nski}
\affiliation{Institute of Physics, Nicolaus Copernicus University \\
Grudzi\c{a}dzka 5/7, 87--100 Toru\'n, Poland}
\author{Paolo Facchi}
\affiliation{Dipartimento di Matematica and MECENAS, Universit\`a di Bari,
        I-70125  Bari, Italy}
\affiliation{INFN, Sezione di Bari, I-70126 Bari, Italy}
\author{Giuseppe Marmo} \affiliation{Dipartimento di Scienze Fisiche  and MECENAS,
Universit\`a di Napoli ``Federico II", I-80126  Napoli, Italy}
\affiliation{INFN, Sezione di Napoli, I-80126  Napoli, Italy}
\author{Saverio Pascazio} \affiliation{Dipartimento di Fisica and MECENAS,
Universit\`a di Bari,
        I-70126  Bari, Italy}
\affiliation{INFN, Sezione di Bari, I-70126 Bari, Italy}

\date{\today}

\begin{abstract}
A time-dependent product is introduced between the observables of a dissipative quantum system, that accounts for the effects of dissipation on observables and commutators. In the $t \to \infty$ limit this yields a contracted algebra. 
The general ideas are corroborated by a few explicit examples.
\end{abstract}

\pacs{03.65.Yz	
}

\maketitle

One of the most distinctive traits of quantum mechanics is the non-commutativity of some of its observables. 
If a commutator vanishes, the associated observables can be simultaneously measured and can be 
considered ``classical" with respect to each other. The system is classical when \emph{all} its observables 
commute. The transition from quantum to classical is a fascinating subject of investigation and interesting 
approaches have been proposed in order to emphasize the role of observables in this context and give a 
consistent definition of classicality \cite{wigner,araki,jauch}.

A dissipative quantum system loses some of its genuine quantum features (such as the ability to interfere) and eventually displays a ``classical" behavior
 \cite{Weiss,Breuer}. 
In this Letter we suggest a mechanism that yields classicality (in the afore-mentioned sense) starting 
from dissipative dynamics and the physics of open quantum systems. Besides 
being of interest in themelves, these subjects have profound conceptual consequences and lead to applications, for example in quantum enhanced applications and quantum technologies \cite{QIT}. It is therefore of interest to understand what happens to the observables of a dissipative quantum system and in which sense measurements yield less information at the end of a dissipative process.
The approach we shall propose is general, but for the sake of simplicity we shall limit our discussion to the 
master equation. Generalizations and further discussion will be postponed to a forthcoming publication.

The description of quantum systems makes use of states $\rho$ and an algebra $\cA$ of observables $A$. 
One can describe the dynamical evolution in terms of the former or the latter, the two pictures being 
equivalent, according to Dirac's prescription \cite{Diracbook}
\begin{equation}
\label{SvsH}
\mbox{Tr}(\rho_t A_0) = \mbox{Tr}(\rho_0 A_t).
\end{equation}
We shall work in the Markovian approximation, when the dynamics
is governed by the master equation
\begin{equation}
\label{M}
    \dot{\rho}_t = L \rho_t ,
\end{equation}
where $\rho_t$ is the density matrix of the quantum system, the subscript $_t$ denotes the evolved quantity at time $t$ and
$L$ is the time-independent generator of a dynamical semigroup. Equation (\ref{M})
can be formally solved
\begin{equation}
\label{Msol}
    \rho_t  = e^{tL}\rho_0 = \Lambda_t (\rho_0) \qquad (t \geq 0)
\end{equation}
and it is well known that under certain conditions on $L$ \cite{GKSL}
the dynamics $\Lambda_t$ is completely positive and trace preserving
\cite{Alicki,Breuer}.

Equation (\ref{SvsH}) leads to the (adjoint) evolution equation for observables (Heisenberg picture)
\begin{equation}
\label{M2}
    \dot{A}_t = L^\sharp A_t  \quad \Leftrightarrow \quad A_t =  \Lambda_t^\sharp (A_0).
\end{equation}
In this Letter we address the following question: what can be meaningfully observed in a dissipative 
quantum system, in particular when it has reached its equilibrium state? Our strategy will be to interpret the 
effects of the adjoint evolution $\Lambda^\sharp$ on the commutators of the algebra of observables $\cA$, 
with basis $\{A_j\}$, defined through its structure constants $C$:
\begin{equation}
\label{Cstruct}
[A_i,A_j] = C^k_{ij} A_k.
\end{equation}
We shall see that in general, the above question will lead to a \emph{contraction} of the algebra of 
observables \cite{IW,Sa}.

\paragraph{First example and general ideas.}
Let us start from a simple but interesting case study. Let
\begin{equation}
\label{qubit3}
   L \rho = - \frac \gamma 2 (\rho- \sigma_3 \rho \sigma_3),
\end{equation}
where $\sigma_\alpha \; (\alpha = 0,1,2,3)$ are the Pauli matrices $(\sigma_0= \openone)$, and $\gamma>0$. This describes the dissipative dynamics of a qubit undergoing phase damping. The asymptotic solution is
\begin{equation}
\label{qubit3sol}
   \rho_0 = \frac{1}{2} (\openone + \bm x \cdot \bm\sigma) \stackrel{t \to \infty}{\longrightarrow}  \Lambda_\infty(\rho_0)=\rho_\infty = \frac{1}{2} (\openone + x_3 \sigma_3),
\end{equation}
$\bm x$ being a vector in the unit 3-dim ball, $|\bm x| \leq 1$.
It is very simple to see that Eq.\ (\ref{qubit3}) yields
\begin{eqnarray}
& & \Lambda^\sharp_t (\sigma_{0,3}) = \Lambda^\sharp_\infty (\sigma_{0,3}) = \sigma_{0,3},
\label{commasympt0} \\
& & \Lambda^\sharp_t (\sigma_{1,2}) = e^{-\gamma t} \sigma_{1,2} \quad \to \quad \Lambda^\sharp_\infty (\sigma_{1,2})=0.
\label{commasympt}
\end{eqnarray}
These equations must be understood in the weak sense, according to Eq.\ (\ref{SvsH}): for example,
the expectation value of $\sigma_{1,2}$ in the asymptotic state (\ref{qubit3sol}) vanishes.
This result offers a remarkable interpretation: as time goes by, it becomes increasingly difficult to measure the coherence between the two states of the qubit. In the $t\to\infty$ limit, coherence is lost and the only nontrivial observables are populations.
This interpretation, although suggestive, must face a serious problem: can one consistentily define a novel product among observables, in such a way that
\begin{equation}
\label{Masympt}
\cA_\infty =  \Lambda^\sharp_\infty (\cA) = \lim_{t \to \infty} \Lambda^\sharp_t (\cA).
\end{equation}
be a well-defined algebra?
The following theorem \cite{CGM} helps answering this question.

Let $\cA$ be a complex topological algebra, i.e., a topological vector
space over $\C$  with  a  continuous  bilinear
operation
\beq
(X,Y) \in \cA\times\cA \mapsto X \cdot Y \in \cA
\eeq
and $U_\lambda:\cA\to\cA$ a family of linear morphisms that continuously
depends on a real parameter $\lambda$.
If $U_\lambda$ are invertible in a neighborhood of the origin $\lambda\in I \setminus\{ 0\}$, then we
can consider the continuous family of products \beq
X\cdot_\lambda Y=U_\lambda^{-1}(U_\lambda(X)\cdot U_\lambda(Y)),
\eeq
for $\lambda\in I \setminus\{ 0\}$. All these products are  isomorphic
by definition, since $U_\lambda(X\cdot_\lambda Y)=U_\lambda(X)\cdot U_\lambda(Y)$
and if $U_0$ is invertible, then clearly
\beq
\label{limN}
\lim_{\lambda\to 0}X \cdot_\lambda Y=U_0^{-1}(U_0(X)\cdot U_0(Y)).
\eeq
However, the $\lim_{\lambda\to 0}X\cdot_\lambda Y$  may  exist  for  all
$X,Y\in\cA$ even if $U_0$ is not invertible and  (\ref{limN})  does  not  make
sense. We say then that $\lim_{\lambda\to 0}X\cdot_\lambda Y$ is a {\it contraction}
of the product $X\cdot Y$. The existence and the form of the contracted product heavily depends on the family $U_\lambda$ \cite{Sa}.

We therefore identify $\lambda = 1/t$, $U_\lambda = \Lambda_t^\sharp$
and adopt the prescription
\begin{equation}
\label{prodpauli}
A \cdot_t B \equiv (\Lambda_t^\sharp)^{-1} (\Lambda_t^\sharp (A) \cdot \Lambda_t^\sharp (B)), \quad \forall  A,B \in \cA .
\end{equation}
Clearly, $ \Lambda_\infty (=U_0)$ is not invertible, but the limiting product ``$\cdot_\infty$" makes sense.
Having defined a  product, we can now define the commutators according to the rule
\begin{equation}
\label{commpauli}
[A_i , A_j]_t \equiv (\Lambda_t^\sharp)^{-1} [\Lambda_t^\sharp (A_i) , \Lambda_t^\sharp (A_j)] \equiv   C^k_{ij}(t) A_k ,
\end{equation}
where $[A,B] = A\cdot B - B \cdot A$.
In the $t \to \infty$ limit Eq.\ (\ref{commpauli})
yields a contraction of the original algebra (\ref{Cstruct})
\footnote{One can also take $U_\lambda = (\Lambda_t^{\sharp})^{-1}$ and define
$A \cdot B \equiv (\Lambda_t^\sharp)^{-1} (\Lambda_t^\sharp (A) \cdot_t 
\Lambda_t^\sharp (B))$, that preserves the product and the 
commutators for any invertible evolution. This definition is also mathematically consistent, but does not yield the same physical interpretation as (\ref{prodpauli}).}.

For instance, in the simple model (\ref{qubit3})-(\ref{commasympt}), the contracted algebra is the Lie algebra of the Euclidean group  $E(2)$ of isometries of the plane:
\begin{eqnarray}
\label{Masympt1}
& & [\sigma_1, \sigma_2]_t = 2 i  e^{-2\gamma t} \sigma_3
 \to 
[\sigma_1,\sigma_2]_\infty=0, \\
& & [\sigma_2, \sigma_3]_t = 2 i  \sigma_1 \to 
[\sigma_2,\sigma_3]_\infty= 2 i  \sigma_1,
\\
& & [\sigma_1, \sigma_3]_t = -2 i  \sigma_2 \to 
[\sigma_1,\sigma_3]_\infty= -2 i  \sigma_2.
\end{eqnarray}
If one adds to $(\ref{qubit3})$ a unitary evolution $-i [H,\rho]$, with Hamiltonian $H=\Omega \sigma_3$, 
nothing changes. However, a Hamiltonian
$H=\Omega \sigma_1$ yields a more involved dynamics \cite{NP99} and makes
$\Lambda^\sharp_\infty (\sigma_{3})$ vanish as well: in this case the contracted algebra is Abelian and
even measurement of populations become trivial. The interpretation is straightforward: the Hamiltonian provokes Rabi oscillations between the two levels, the asymptotic state is
$\rho_\infty = \openone /2$ [rather than (\ref{qubit3sol})] and the final state is totally mixed.
Having tested our general scheme on a simple but significant example, we can now look at more complicated situations.

\paragraph{Second example.}
Let
\begin{equation}
\label{oscdamp}
   L \rho =  -\frac{\gamma}{2} \left( \{a^\dagger a,\rho \} -2 a\rho a^\dagger \right),
\end{equation}
that describes a harmonic oscillator undergoing energy damping. Here, $\{A,B\}=AB+BA$. It is easy to check that
\begin{eqnarray}
\label{oscdampt2}
& & \Lambda^\sharp_t (a) = e^{-\gamma t/2} a , \quad \Lambda^\sharp_t (a^\dagger) = e^{-\gamma t/2} a^\dagger , \nonumber \\
& & \Lambda^\sharp_t (N) = e^{-\gamma t} N \quad (N=a^\dagger a),
\end{eqnarray}
so that the oscillator algebra is contracted to an Abelian algebra, with $[a,a^\dagger]_\infty=[a,N]_\infty=0$ 
(remember that the above equations are understood in the weak sense). The 
physical picture is straightforward: dissipation drives the system to its ground state and in the limit
not only the relative coherence, but even the populations of the excited states vanish.
The introduction of a Hamiltonian $H=\omega a^\dagger a$ does not change the global picture.

\paragraph{Third example.}
Let
\begin{equation}
\label{oscphdamp}
   L \rho =  -\frac{\gamma}{2} \left(\{(a^\dagger a)^2,\rho \} -2 a^\dagger a\rho a^\dagger a \right),
\end{equation}
that describes a harmonic oscillator undergoing phase damping. Since $L^\sharp=L$ and $\Lambda^\sharp=\Lambda$, one finds
\begin{eqnarray}
\label{oscphdampt2}
& & \Lambda^\sharp_t (a) = e^{-\gamma t/2} a , \quad \Lambda^\sharp_t (a^\dagger) = e^{-\gamma t/2} a^\dagger , \nonumber \\
& & \Lambda^\sharp_t (N) = N ,
\end{eqnarray}
so that, unlike in the second example, $N$ is left unaltered. 
The contraction of the oscillator algebra yields the Lie algebra of the Poincar\'e group in 1+1 dimensions ISO(1,1):
\begin{equation}
[a,a^\dagger]_{\infty} =  0, \qquad [a,N]_{\infty}   =  a, \qquad [a^\dagger, N]_{\infty}  =-a^\dagger .
\end{equation} 
The physical picture is straightforward: in the presence of phase damping the system is driven to an 
incoherent mixture (in the energy basis). However, in the asymptotic limit it is still possible to measure 
nonvanishing populations of the different states. 
The introduction of a Hamiltonian $H=\omega a^\dagger a$ does not change anything.

\paragraph{Fourth example.}
Let
\begin{equation}
\label{pq}
   L \rho =  -\gamma(\{x^2,\rho \} -2 x\rho x ) = - \gamma [x,[x,\rho]],
\end{equation}
that describes a massive particle undergoing decoherence:
\begin{equation}
\label{offvanishcont}
    L |x\>\<y| = - \gamma (x-y)^2 |x\>\<y| .
\end{equation}
Also in this case, the generator (\ref{pq}) is self-dual, $L=L^\sharp$.

By considering formally $x$ and $p$ as bounded operators, one gets
\begin{equation}
\label{pqt}
\Lambda^\sharp_t (p) = p , \qquad \Lambda^\sharp_t (x) = x,
\end{equation}
for all $t$,
so that the CCR are preserved. 
However one gets, for $n\geq 2$,
\begin{equation}
L(p^n) = \gamma n (n-1) p^{n-2},
\end{equation}
so higher order commutation relations change.

These findings can be corroborated by working with the (bounded) unitary groups generated by $x$ and $p$, that is
the Weyl operators
\begin{equation}\label{}
    U(\alpha) = e^{i \alpha x}, \quad V(\beta) = e^{i\beta p},
    \quad \alpha, \beta \in \mathbb{R} .
\end{equation}
They satisfy
\begin{equation}\label{}
    U(\alpha)V(\beta) = e^{-i\alpha\beta} V(\beta)U(\alpha) .
\end{equation}
One has $[x,U(\alpha)]=0$ and $[x,V(\beta)] = - \beta V(\beta)$, yielding
\begin{equation}
\label{}
    L\, U(\alpha) = 0, \quad   L\,V(\beta) = - \gamma \beta^2\, V(\beta) ,
\end{equation}
and hence
\begin{equation}\label{LambaWeyl}
    \Lambda^\sharp_t U(\alpha) = U(\alpha),  \quad   \Lambda^\sharp_t V(\beta) = e^{-\gamma \beta^2 t}
    V(\beta).
\end{equation}
Notice, that for any $\beta\neq0$ $\Lambda^\sharp_t V(\beta)$ is no longer unitary for $t>0$, and 
asymptotically vanishes. However, for any $t$ one has
\begin{eqnarray}
 U(\alpha)\cdot_t V(\beta) =e^{-i\alpha\beta} V(\beta)\cdot_t U(\alpha) ,
\end{eqnarray} 
that is, the commutation relations of the Weyl system are preserved. However, the Weyl system itself is not 
preserved, since $\Lambda^\sharp_t V(\beta)$ is not unitary. This example clarifies that, while the contraction 
does not affect the basic Lie algebra, it changes the  \emph{whole} associative algebra, and thus the 
higher-order commutators.
Finally, notice that the presence of a free Hamiltonian changes the picture considerably \cite{venugopalan} 
and will not be considered here.

\paragraph{Fifth example.} Finite dimensional version of the fourth example.
Consider a $d$-level system and let
\begin{equation}
X = \sum_{m=1}^{d} m
|m\>\<m|
\end{equation}
be the discrete position operator on a circle. Consider the analogous of (\ref{pq})
\begin{equation}\label{XX}
    L\rho = - \gamma [X,[X,\rho]]. 
\end{equation}
Let us introduce Schwinger's  unitary operators \cite{Schwinger}
\begin{equation}
    U = \sum_{m=1}^d \lambda^m |m\>\<m|  , \quad  V= \sum_{k=1}^d \lambda^{-k}
    |\widetilde{k}\>\<\widetilde{k}|,
\end{equation}
where $\lambda=e^{2\pi i/d}$, and the momentum
eigenbasis $\{|\widetilde{k}\>\}$, defined by a discrete Fourier transform,
\begin{equation}
    |\widetilde{k}\> = \frac{1}{\sqrt{d}} \sum_{m=1}^{d} \lambda^{-km} |m\> .
\end{equation}
Schwinger's system, which is the finite dimensional version of Weyl's, satisfies
\begin{equation}
    U^k V^l = \lambda^{kl} V^l U^k ,
\end{equation}
for $k, l=1,\ldots,d$.
One easily finds [compare with (\ref{LambaWeyl})]
\begin{equation}
    \Lambda^\sharp_t U^k = U^k\ , \ \ \ \   \Lambda^\sharp_t V^l = e^{-\gamma l^2 t}
    V^l ,
\end{equation}
so that $V^l$ asymptotically vanishes. Again, $\Lambda^\sharp_t V^l$ is no longer unitary for $t>0$.
 As a consequence, like in the previous example, we get
\begin{equation}
\label{}
U^{k} \cdot_t V^{l} = \lambda^{kl} V^{l} \cdot_t U^{k} ,
\end{equation}
and the commutation relations are preserved. However, Schwinger's system
is not preserved, since $\Lambda^\sharp_t V^l$ is no longer unitary.  From Eq.~(\ref {XX}) one has the discrete version of
(\ref{offvanishcont})
\begin{equation}
\label{offvanish}
    L |m\>\<n| = - \gamma (m-n)^2 |m\>\<n| ,
\end{equation}
so that each observable becomes asymptotically 
diagonal in the position eigenbasis $|m\>$.
It is clear that the introduction of a unitary evolution with Hamiltonian $H=\sum_m h_m
|m\>\<m|$ does not change the global picture.

\paragraph{Sixth example.}
Finally, let us consider the following model of pure decoherence of a
$d$-level system. Define $d$ unitary operators
\begin{equation}\label{Uk}
    U_k = \sum_{l=0}^{d-1} \lambda^{-kl} P_l ,
\end{equation}
where $P_l=|l\>\<l|$ and $\lambda=e^{2\pi i/d}$.  Note that $U_0 =
\openone_d$, and ${\rm Tr}\, U_k = 0$ for $k\geq 1$. Now, for
$\gamma_1,\ldots,\gamma_{d-1}\geq 0$ let us define the following
generator
\begin{equation}\label{L-d}
    L\rho = - \frac 1d\, \sum_{k=1}^{d-1} \gamma_k \left( \rho - U_k \rho U_k^\dagger \right) .
\end{equation}
It is clear that for $d=2$ one has $U_1 = \sigma_3$ and hence
(\ref{L-d}) reproduces (\ref{qubit3}) as a particular case. Using (\ref{Uk}) one easily
derives the dynamical map
\begin{equation}\label{}
    \Lambda_t \rho = \sum_{m,n=0}^{d-1} c_{mn}(t) P_m \rho P_n ,
\end{equation}
where the decoherence matrix $c_{mn}(t)$ reads
\begin{eqnarray}
\label{}
 c_{mn}(t) &=& e^{-(i\omega_{mn} + \gamma_{mn})t} , \\
     \gamma_{mn} &=& \frac 1d \sum_{k=1}^{d-1} \gamma_k \, {\rm Re} \left(1- \lambda^{-k(m-n)}\right) , \nonumber \\
   \omega_{mn} &=&  - {\rm Im}\, \left(\sum_{k=1}^{d-1} \gamma_k \lambda^{-k(m-n)}  \right). \nonumber 
\end{eqnarray}
Note that $\gamma_{mn} = \gamma_{nm}$, with $\gamma_{mm}=0$, and 
$\omega_{mn} = -\omega_{nm}$, which implies
$\omega_{mm}=0$. In particular, if all
$\gamma_j=\gamma$, then
\begin{equation}\label{}
    \gamma_{mn}=\gamma\ \ (m\neq n), \ \ \ \ \omega_{mn}=0 ,
\end{equation}
and one finds
\begin{equation}\label{}
    \Lambda_t^\sharp |m\>\<n| = c_{nm}(t)\, |m\>\<n| .
\end{equation}
Hence, due to $\gamma_{mn} >0$, only the diagonal elements $P_m$
survive asymptotically. If one adds to (\ref{L-d}) the Hamiltonian
$H = \sum_k h_k P_k$, the asymptotic picture does not
change. Finally, one finds the following formula for the product $A \cdot_t B$
\begin{equation}\label{}
    |m\>\<n| \cdot_t |k\>\<l| = \frac{c_{nm}(t)
    c_{lk}(t)}{c_{lm}(t)}\,  |m\>\<n| \cdot |k\>\<l|\ .
\end{equation}
In particular, if all decoherence rates are equal
$\gamma_j=\gamma$, 
\begin{equation}\label{}
    |m\>\<n| \cdot_t |k\>\<l| = e^{-\gamma [1 + \delta_{ml} - \delta_{mn} - \delta_{kl} ]t} \,\delta_{nk}\,   |m\>\<l|
    .
\end{equation}

\paragraph{Conclusions.} Starting from the adjoint evolution of a dissipative quantum system, we have 
defined a product that yields a contracted algebra of observables. Other definitions, fully consistent from a 
mathematical point of view, are clearly possible, but do not yield an equally appealing physical 
interpretation. In some sense, the ansatz 
(\ref{prodpauli}) ``ascribes" to the product $\cdot_t$ the dissipative features of the evolution and the 
increasing difficulty in measuring those observables that are more affected by decoherence and dissipation. 

In the present framework, ample room is left for noncommutative (quantum) observables, that do not 
belong to the center of the contracted algebra. These are associated with the kernel 
of $L^\sharp$. These observables are not affected by dissipation and preserve their quantum features. One 
can find many examples, e.g.\ in models like those discussed in the sixth example (when some $\gamma_
{mn}=0$).

We confined our analysis to Markovian systems, described by the master equation  (\ref{M}). However, our 
main conclusions remain valid when the evolution is described by a map (quantum channel). This unearths 
additional possibilities that will be discussed in a forthcoming paper.

\acknowledgments The authors thank
SVYASA University, Bangalore (India) for their warm hospitality during the final part of this work.

\end{document}